\documentclass[prl,amsmath,amssymb,lengthcheck,showpacs,floatfix,groupedaddress,superscriptaddress,longbibliography]{revtex4-1}
\usepackage{amsmath}
\usepackage{amsfonts}
\usepackage{amssymb}
\usepackage{graphicx}
\usepackage[english]{babel}
\usepackage{times}
\usepackage{dcolumn}
\usepackage[usenames,dvipsnames]{color}
\usepackage{units}
\usepackage[utf8]{inputenc}
\usepackage{hyperref}
\hypersetup{
  colorlinks   = true,    
  urlcolor     = blue,    
  linkcolor    = blue,    
  citecolor    = green      
}

\begin{document}

\title{Universal magnetic oscillations of DC conductivity in the incoherent
regime of correlated systems}

\author{Jakša Vučičević}

\affiliation{Scientific Computing Laboratory, Center for the Study of
Complex Systems,\\
Institute of Physics Belgrade, University of Belgrade, Pregrevica 118,
11080 Belgrade, Serbia}

\author{Rok \v{Z}itko}

\affiliation{Jo\v{z}ef Stefan Institute, Jamova 39, SI-1000 Ljubljana, Slovenia}
\affiliation{Faculty of Mathematics and Physics, University of Ljubljana,
Jadranska 19, SI-1000 Ljubljana, Slovenia}

\date{\today}

\begin{abstract}
Using the dynamical mean field theory 
we investigate the magnetic field dependence of DC conductivity in the
Hubbard model on the square lattice,
fully taking into account the orbital effects of the field introduced
via the Peierls substitution.
In addition to the conventional Shubnikov-de Haas 
quantum
oscillations, associated with the coherent cyclotron motion of
quasiparticles and the presence of a well-defined Fermi surface, we
find an additional oscillatory component with a higher frequency that
corresponds to the total area of the Brillouin zone.  
These
paradigm-breaking oscillations appear
at elevated temperature.
This finding is in excellent qualitative agreement with the recent
experiments on graphene superlattices.
We elucidate the key roles of the off-diagonal elements of
the current vertex and the incoherence of electronic states, and explain the
trends with respect to temperature and doping.
\end{abstract}


\maketitle

\newcommand{\expv}[1]{\langle #1 \rangle}
\newcommand{\tk}{\tilde{\mathbf{k}}}
\newcommand{\ImG}{\mathrm{Im}G}

Quantum oscillations (QOs) are a fundamental phenomenon in solid state physics.
The Lorentz force affects electrons in such a way that all the system properties
vary periodically with the inverse of the magnetic field\cite{shoenberg}.
Conventionally, QOs are observable at low temperatures $T$ and in absence of strong incoherence,
and provide detailed information about the topology and shape of the Fermi surface.\cite{shoenberg,LK}
Yet, QOs are surprisingly ubiquitous. They also appear in non-Fermi liquids
\cite{Denef2009,Hartnoll2010,Else2020} and even in gapped systems such
as Kondo insulators \cite{Knolle2015}. They were observed in graphite
\cite{Soule1964,Hubbard2011}, graphene \cite{Novoselov2005,Zhang2005},
organics \cite{Kartsovnik2005}, cuprates
\cite{DoironLeyraud2007,Sebastian2008,Sebastian2015}, perovskite
heterostructures \cite{Caviglia2010,Moetakef2012}, iron-pnictide
superconductors \cite{Carrington2011}, and moir\'e
systems \cite{Cao2016}.

In moir\'e systems, huge superlattice spacing allows access
to regime of large flux per unit cell $\Phi$.
Precisely in this regime, 
recent experiments\cite{Hunt2013,KrishnaKumar2017,KrishnaKumar2018,Barrier2020} 
have uncovered
a new, peculiar type of QOs of conductivity:
peaks
at $\Phi$ equal to simple fractions of the flux quantum, i.e.
$\Phi=\Phi_0 p/q $ with $p$, $q$ coprime integers, and $p$ and $q$ small\cite{KrishnaKumar2018}.
These Brown-Zak (BZ) oscillations are clearly distinct from the 
conventional Shubnikov-de Haas (SdH) oscillations:
BZ QOs appear at elevated temperatures\cite{KrishnaKumar2017}, and their frequency does not depend on the electron density $n$ 
(in 2D, SdH QOs have a frequency proportional to $n$).
Some understanding of this phenomenon was reached by noting that the conductivity is high whenever the non-interacting density of states (DOS) consists of 
a small number ($q$) of wide energy bands (magnetic ``minibands'')\cite{KrishnaKumar2017,KrishnaKumar2018}.
States in wider bands should have a higher velocity, and therefore conduct better.
However, this heuristic picture cannot explain the totality of experimental observations.
In this paper we present a microscopic theory of conductivity
in the Hubbard model and unexpectedly recover a phenomenology
strikingly similar to that observed in the experiments of
Refs.~\onlinecite{KrishnaKumar2017} and \onlinecite{KrishnaKumar2018}.
Our analysis elucidates the essential role of incoherence for the BZ oscillations, and explains the temperature, doping and interaction
trends in a systematic manner.

We employ the recently developed extension of the dynamical mean field theory (DMFT)\cite{georges1996} to finite magnetic fields
\cite{Acheche2017,Markov2019,OURPRB}.
In absence of the magnetic field, the DMFT solution of the Hubbard model was previously shown to describe the transport properties of various materials\cite{Limelette2003,Terletska2011,vucicevic2013,Furukawa2015,vucicevic2015prl} 
and cold atoms in optical lattices\cite{Vucicevic2019,Brown2018}.
The DMFT approximates the self-energy by a local quantity, and becomes exact in the limit of infinite coordination number.
In a separate accompanying publication Ref.~\onlinecite{OURPRB}, we prove that the vertex corrections for the longitudinal conductivity cancel at the level of DMFT, regardless of the magnetic field (see also Refs.~\onlinecite{khurana1990} and \onlinecite{Markov2019}); this makes it possible to calculate conductivity by the Kubo bubble without any additional approximations.
Our approach fully takes into account local correlations due to
electron-electron (e-e) interaction, and is formally applicable at any
$T$, coupling strength $U$ and field $B$.

\begin{figure*}
\centering
\includegraphics[width=2.0\columnwidth, trim=0cm 0 0 0, clip]{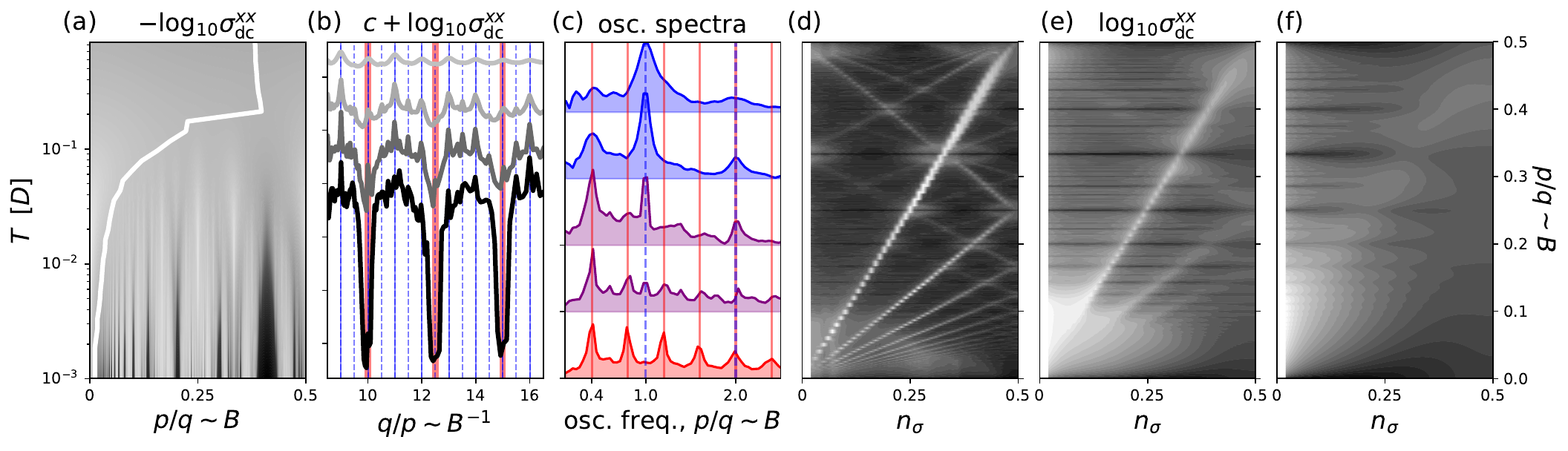}
\caption{ DMFT results for conductivity in the Hubbard model for $U=1D$. 
(a) Conductivity as a function of temperature and field at band filling $n_\sigma=0.4$.
Colorcode is logarithmic: black means $\log_{10} \sigma^{xx}_\mathrm{dc}\approx-7.95$, white means $\log_{10} \sigma^{xx}_\mathrm{dc}\approx2.12$.
White line: onset points of the non-monotonic behavior of $\sigma^{xx}_\mathrm{dc}(B)|_T$.
(b) Conductivity as a function of inverse magnetic field. Bottom to top: $T=0.0012, 0.0049, 0.0109, 0.024 D$; 
lines are plotted on the log scale, and offset for the sake of clarity.
(c) Frequency spectrum of conductivity in the range $p/q\in[0.03,0.15]$ at different temperatures. Bottom to top: $T=0.001, 0.009,0.016, 0.029, 0.064 D$.
Each spectrum is normalized to 1 and shifted for the sake of clarity.
(b,c) Verical lines: peaks due to SdH oscillations (red), and BZ oscillations (blue).
(d,e,f) Conductivity with respect to band filling and field at $T=0.005,0.03,0.1D$, respectively.
Colorcode: white means -8.22, -3.57, -3.12, black means
2.14, 1.77, 1.03, respectively.
}
%
\label{fig1}
\end{figure*}

Our conductivity results exhibit oscillations
that clearly correspond to the BZ QOs observed in experiment.
The oscillations have a frequency $p/q=1$ (corresponding to maxima at $p/q=1/q$)
and appear at relatively high $T$ where the SdH oscillations are getting
thermally washed out.
BZ either coexist with the SdH oscillations or appear as the sole oscillatory component.
As $T$ is lowered, higher harmonics of BZ oscillations become more
pronounced (peaks become sharper, and additional maxima at $p/q=2/q,3/q...$ appear). Ultimately, at very low $T$,
regular BZ oscillations give way to fractal behavior which does not yield any pronounced peaks in the Fourier spectrum.
It turns out that the essential ingredient for the regular (sinusoidal) BZ oscillations are the
incoherent electronic states.
Incoherence allows for conduction processes that involve tunneling
between two eigenstates of the Hamiltonian,
and it is precisely the contribution of those procesess that oscilates
at frequency $p/q=1$.
Our numerical data suggests that in strongly correlated regimes,
regular BZ oscillations should appear at very low temperature.

\emph{Model and method.} We consider the Hubbard model on the square
lattice with nearest-neighbor hopping $t$,
%
coupling $U$ and band filling per spin $n_\sigma$, with $n = \sum_\sigma n_\sigma$.
We use $D=4t$ as the unit of energy.
The field is included through Peierls phases for rational flux values $\Phi/\Phi_0 = p/q$ to obtain commensurate magnetic cell \cite{Hofstadter1976}. 
We do not include the Zeeman term
\cite{laloux1994,bauer2007fm}, as it 
does not affect the QO
frequencies, only their amplitudes\cite{shoenberg}. 
We solve the problem within the DMFT with numerical renormalization group
solver.
Full details of our calculations are given in Ref.~\onlinecite{OURPRB}.

\emph{Results.} 
Fig.~\ref{fig1}(a) shows the conductivity 
for moderate doping and interaction ($n_\sigma =0.4$, $U=1$)
over a broad range of temperature and field (flux). At low $T$, we clearly see
prominent oscillations.
The onset of non-monotonic behavior is marked with the white line:
it indicates the value of $B$ where the first extremum in $\sigma^{xx}_\mathrm{dc}$ is encountered
for a given $T$.
On Fig.~\ref{fig1}(b) we zoom in on a narrow field range and plot
$\sigma_\mathrm{dc}^{xx}$ as a function of $1/B$ at several $T$.
At low $T$, we see large dips in conductivity for $p/q =
n_\sigma/i$ (red lines; $i$ is integer), corresponding to occurrences
of a large gap in the density of states at the Fermi level.
These are the SdH oscillations with a frequency related to the area of the Fermi sea $A_\mathrm{FS}$ by the Onsager relation
$F =\Phi_0/(2\pi)^2 A_\mathrm{FS}$, $A_\mathrm{FS} = (2\pi)^2 n_\sigma
$. 
In between the sharp SdH dips, one can observe a weak but highly
non-monotonous behavior of $\sigma^{xx}_{dc}$ with high-frequency
oscillatory features exceeding the resolution of our calculations.
With increasing $T$, the amplitude of the SdH oscillations is reduced in line with the Lifshitz-Kosevitch theory \cite{LK, OURPRB}, and
the behavior in between the SdH dips becomes simpler: one gets 
spikes coinciding with small-$p$/moderate-$q$ values of flux
(denoted with blue lines: full line is $p=1$, dashed line is
$p=2$). Ultimately, only regular sinusoidal oscillations of period 1
remain, with maxima at $p/q=1/q$. Increasing $T$ further erases all non-monotonic behavior.

Fig.~\ref{fig1}(c) shows the oscillation spectra obtained by Fourier transforming $\sigma_\mathrm{dc}^{xx}(B^{-1}\sim q/p)$
on the range $p/q\in[0.03,0.15]$. 
At the lowest temperature we see strong peaks at $p/q=n_\sigma
$ and its higher harmonics, corresponding to (sharp) SdH oscillations.
The fractal behavior in between the SdH dips seen in Fig.\ref{fig1}(b) does not produce a clear oscillatory signal\cite{OURPRB}.
As $T$ is increased, the peaks at $p/q=1$ and $p/q=2$ appear, while at
the highest $T$ one is left only with the peak at $p/q=1$.

\begin{figure*}
\centering
\includegraphics[width=0.66\columnwidth]{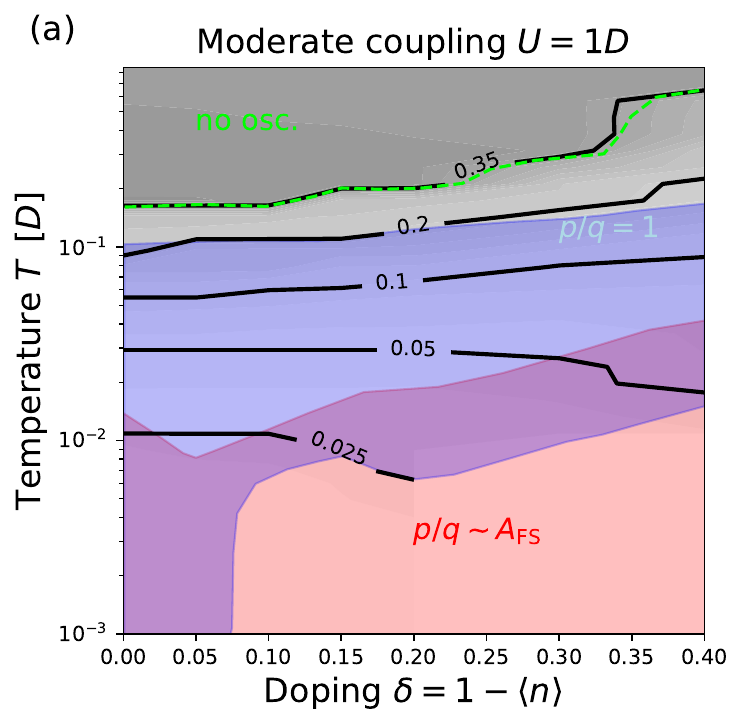}
\includegraphics[width=0.66\columnwidth]{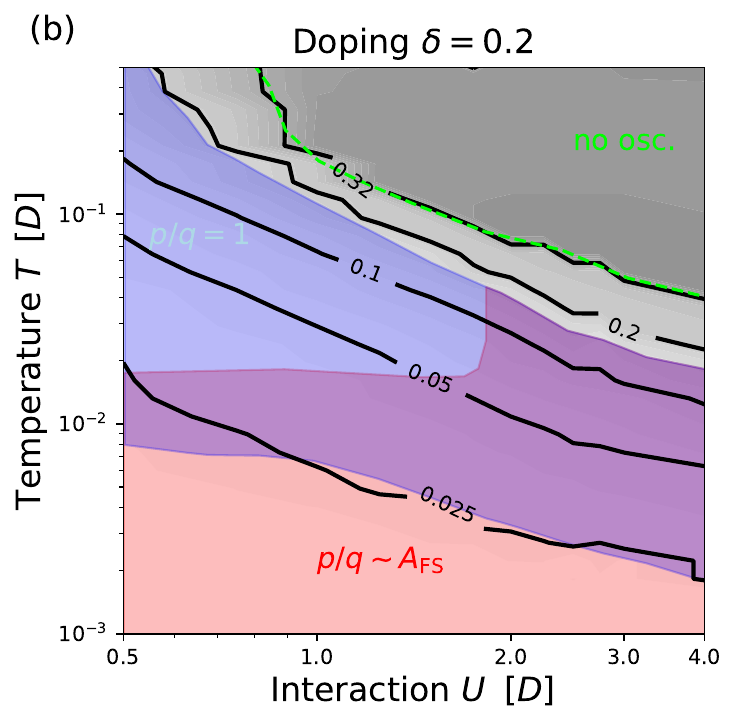}
\includegraphics[width=0.66\columnwidth]{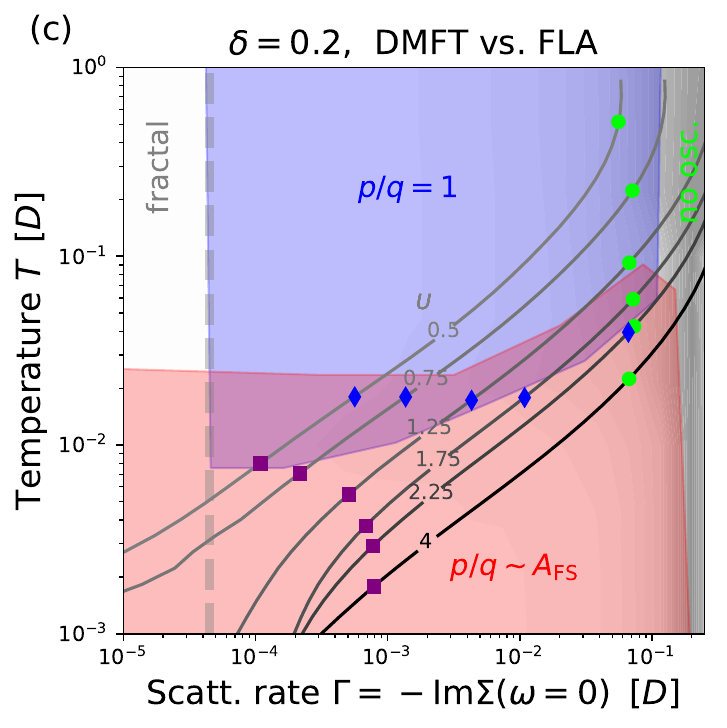}
\caption{
Phase diagrams showing the type of QOs observed
in the range of field $p/q\in [0.03,0.15]$.
(a) DMFT results in $(\delta,T)$ plane, (b) DMFT results in $(U,T)$ plane, (c)
FLA results in $(\Gamma,T)$ plane.
Red: SdH only. Purple: both SdH and BZ, but SdH dominant. Blue: BZ dominant ($p/q=1$
peak stronger than $p/q\approx n_\sigma$ peak).
Black shading and contours in (a,b) denote the value of the field where non-monotonic behavior starts in $1/\sigma_\mathrm{dc}^{xx}(B)|_T$ (analogous to the white line in Fig.\ref{fig1}).
Above the lime dashed line, no oscillations are detectable at any field strength.
In (c), lines and symbols correspond to DMFT results, shading to FLA
results. Lines are $\Gamma(T)$ for various values of $U$. Purple squares indicate where the BZ oscillations start with
increasing $T$, blue diamonds where the BZ becomes dominant, and
lime circles where all QOs cease (corresponding to the top edge of blue and purple regions in (b)).
}
\label{phaseDiagram}
\end{figure*}

On Fig.~\ref{fig1}(d,e,f) we plot the conductivity in the
$(n_\sigma, B)$ plane.
At low $T$, the SdH oscillation fans out from the $(0, 0)$ point, clearly indicating the $n_\sigma$ dependence of the oscillation frequency.
At a higher $T$, SdH oscillations become weaker; horizontal (i.e. $n_\sigma$-independent) stripes corresponding to fractal BZ oscillations become visible, and are particularly
pronounced at small $p$ values.
At the highest $T$ shown, only the BZ oscillations remain.


%


We summarize our observations by presenting in Fig.~\ref{phaseDiagram}(a,b) the two relevant Hubbard model phase diagrams, 
showing the dominant type of (regular) oscillations, based on the Fourier spectrum of $\sigma_\mathrm{dc}^{xx}(q/p)$
in the field range $p/q\in[0.03,0.15]$.
We also indicate the onset field for the non-monotonic behavior (grayscale colorcoding and the black contours).
Clearly, the onset field depends strongly on $U$ and $n$; 
the non-monotonic behavior is stronger and requires less strong fileds in more coherent regimes (lower $U$ and/or higher doping away from half-filling $\delta=1-n$).
Another notable trend is that the BZ oscillations start at a lower
temperature in less coherent regimes (lower $\delta$ at fixed
$U$; stronger $U$ at fixed $\delta$).

To elucidate the role of incoherence 
we perform calculations within the the finite-lifetime approximation (FLA)\cite{OURPRB},
where lifetime of electronic states is set by hand by fixing the (local)
self-energy to $\Sigma(\omega)=-i\Gamma$.
We determine the phase diagram of FLA with respect to the two parameters of this toy model,
the scattering rate $\Gamma$ and tempertaure $T$ (Fig.~\ref{phaseDiagram}(c)). 
There appears to be a well defined upper cutoff value of $\Gamma$ for
the observation of any QOs. For the observation of SdH oscillations, there is a relatively well defined upper cut-off $T$.
The region of dominant regular BZ oscillations is additionally limited by lower cut-off $\Gamma$ and $T$. Below $\Gamma\approx5\times10^{-5}$, fractal behavior is observed, with or without the SdH oscillations, depending on temperature.
At moderate $\Gamma$, increasing the temperature alone does not wash out the BZ oscillations, and they persist up to infinite temperature.

\begin{figure}[h!]
\centering
\includegraphics[width=1.\columnwidth]{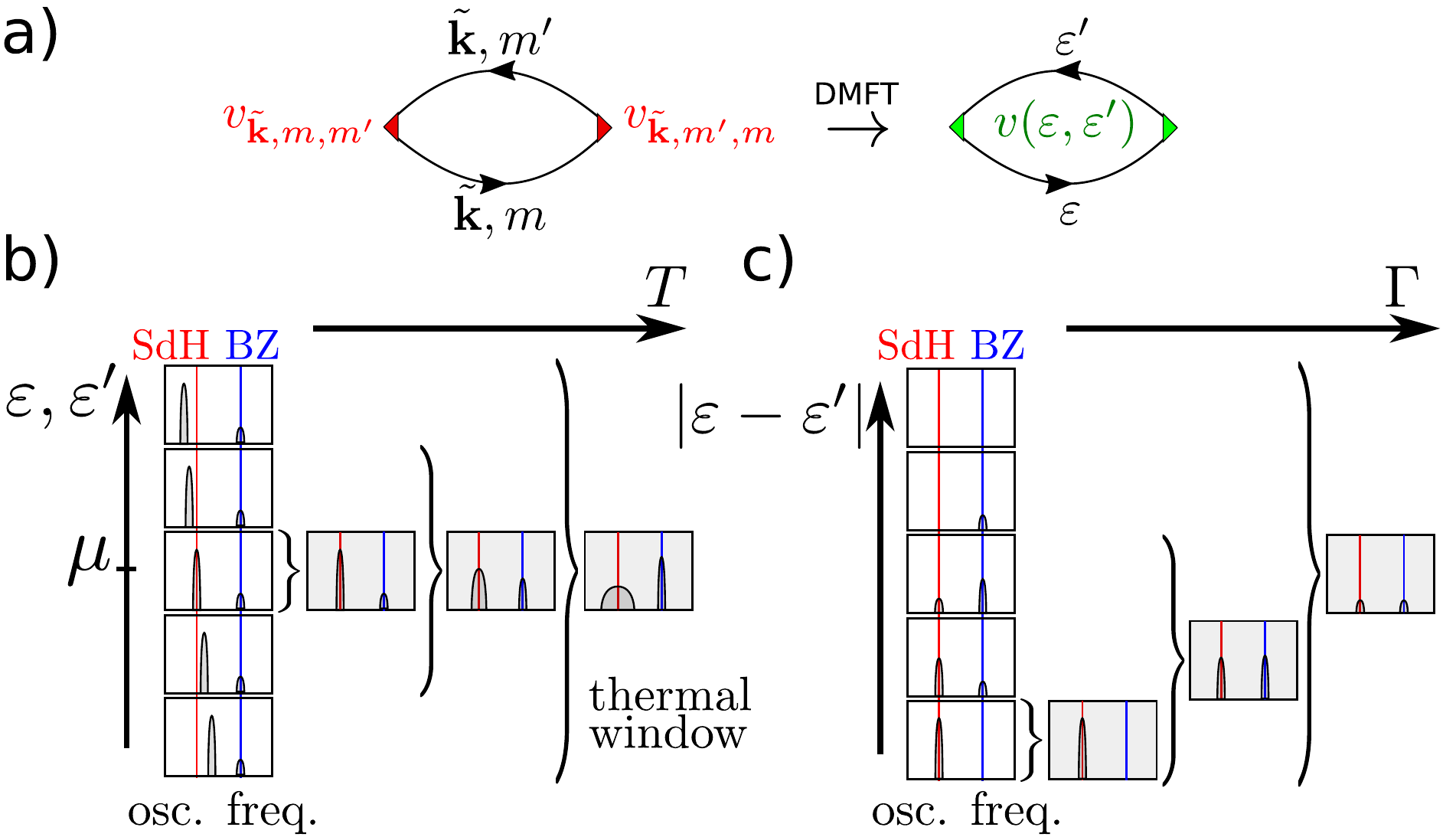}
\includegraphics[width=1.\columnwidth]{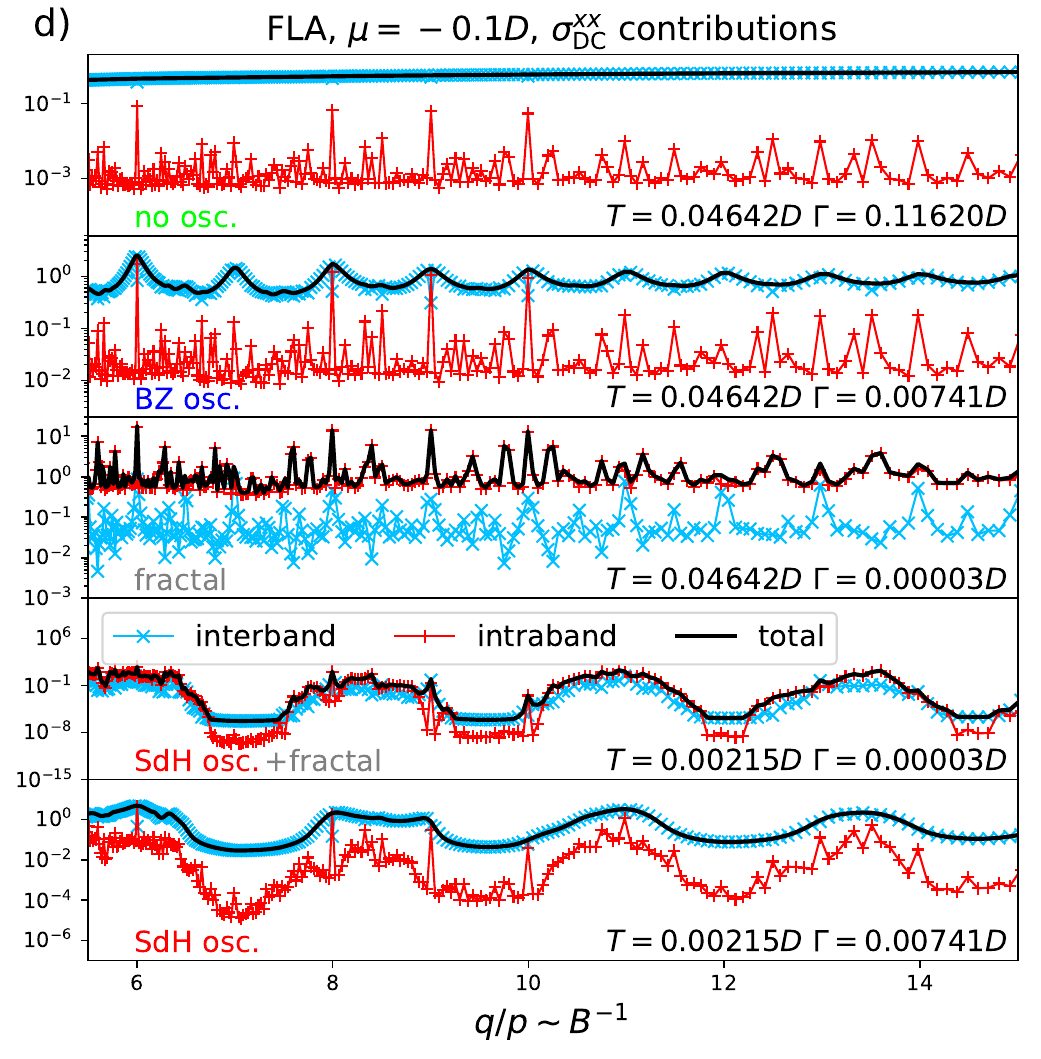}
\caption{(a) Diagrammatic representation of the Kubo bubble. Left: in general; right: at the level of the DMFT.
$(\tilde{\mathbf{k}},m)$ denotes eigenstates of the non-interacting Hamiltonian (see \cite{OURPRB} for details).
Red/lime triangles are the velocity vertices, in DMFT rewritten as a single factor depending on two kinetic energies, 
$v(\varepsilon,\varepsilon')$.
(b,c) White panels: Oscillation spectra of $v(\varepsilon,\varepsilon')$ at a given $(\varepsilon,\varepsilon')$. 
Gray panels: Oscillation spectra of $v$ integrated over the relevant $(\varepsilon,\varepsilon')$-domain, depending on model parameters ($T$ and $\Gamma$), 
as indicated by the large curly bracket; (b) Trend with respect to temperature.
(c) Trend with respect to the scattering rate. 
(d) Field dependence of conductivity and the contributions of interband ($\epsilon\neq \epsilon'$) and intraband ($\epsilon\approx\epsilon'$) processes in FLA in four different parameter regimes.
}
\label{fig:illustration}
\end{figure}

We superimpose on the FLA phase diagram the DMFT results by identifying
$\Gamma = -\mathrm{Im}\Sigma(\omega=0)$. In DMFT the self-energy has
frequency dependence and depends on both $U$ and $T$. The grayscale
lines represent the DMFT result for $\Gamma(T)$ for different $U$
values. The upper cut-off $\Gamma$ for QOs (lime points) holds in good
agreement with FLA results, as well as the upper cut-off $T$ for SdH
oscillations (blue diamonds). At low $U$, the lower cut-off $T$ for BZ
oscillations is also in agreement with FLA. However, at high $U$, the
discrepancy from FLA is significant: the sinusoidal BZ oscillations
appear at much lower $T$ than one would expect based on a simple FLA
toy model where $\Sigma$ has no frequency dependence. At very strong
$U$, there rather seems to be a well defined lower cut-off $\Gamma$
for regular BZ QOs extending to very low $T$ (this lower $\Gamma$ cut-off being a bit higher than the one at high $T$).
The observation of BZ oscillations at very low $T$ is therefore a clear indication of strong electronic correlations that go beyond simple incoherence effects.

\emph{Discussion.} 
The trends related to incoherence and temperature can be understood
from the linear-response transport theory underlying our calculations.
The Kubo bubble for conductivity is illustrated in Fig.~\ref{fig:illustration}(a).
At the level of the DMFT where the self-energy does not depend on the
momentum, the product of two velocities
$v_{\tilde{\mathbf{k}},m,m'}v_{\tilde{\mathbf{k}},m',m}$ can be
rewritten as a single factor with two kinetic-energy
arguments, $v(\epsilon,\epsilon')$.
Depending on temperature, effective scattering rate and chemical potential,
different $(\epsilon,\epsilon')$ domains play a role \cite{OURPRB}.
In particular, only $(\epsilon,\epsilon')$ such that $|\epsilon -\epsilon'|<\Gamma$ 
and $\epsilon^{(\prime)}-\mu<T$ give significant contributions.
At low $T$, we observe that the SdH effect is already contained in
$v(\epsilon,\epsilon')$. 
The oscillation spectrum for $v(\epsilon,\epsilon'\approx\epsilon\approx\mu)$, exhibits
a peak that moves with $\mu$ and coincides with $n_\sigma$.
As the thermal window becomes larger, a wider range of $v(\epsilon,\epsilon'\approx\epsilon)$ enter the calculation,
yet oscillate with different frequencies, depending on $\epsilon$.
This leads to dephasing and washing out of the SdH oscillations.
By contrast, the BZ oscillation is mild at any given $\epsilon$, 
but it \emph{always has the same frequency} ($p/q=1$), thus its
contribution accumulates with increasing $T$ and can become the
dominant effect, as illustrated in Fig.~\ref{fig:illustration}(b).
The domain of $v$ that turns out to oscillate with the BZ frequency 
is found at moderate $|\epsilon-\epsilon'|$.
Therefore, as the scattering rate $\Gamma$ is increased, 
those values enter the calculation and the BZ oscillations become visible in $\sigma^{xx}_\mathrm{dc}(q/p)$.
The values of $v(\epsilon,\epsilon')$ at large $|\epsilon-\epsilon'|$ do not oscillate with any particular frequency. 
As those get included at large $\Gamma$, all oscillations are
ultimately overcome by the non-oscillatory contributions,
as illustrated in Fig.~\ref{fig:illustration}(c).
The velocity $v$ is the only source of BZ oscillations in the Kubo bubble, as the Green's function and the self-energy
do not have an oscillatory component at the frequency of BZ oscillations\cite{OURPRB}.

In previous works\cite{KrishnaKumar2017,KrishnaKumar2018}, the BZ oscillations were connected with
the velocity of the magnetic minibands, calculated as $v=\partial \epsilon_{\tilde{\mathbf{k}},m}/\partial \tilde{k}_x$.
Nevertheless, it is important to note that the
eigenstates of the non-interacting Hamiltonian do not have a well defined velocity in
the presence of the field.
Rather, the velocity $v_{\tilde{\mathbf{k}},m,m'}$ is \emph{a matrix} in the miniband space $m,m'$.
In previous works this was not taken into account
and the results were interpreted
in terms of only the \emph{intraband} processes (diagonal elements of $v$). 
This would be well justified only in the limit of coherent,
long-lived quasiparticle states. However, increasing $T$ even at weak
coupling leads to decoherence of electron states, which activates the
contribution of off-diagonal velocity components and even makes them
fully dominant\cite{OURPRB}.
This corresponds to $m\neq m'$ (or $\varepsilon\neq\varepsilon'$) terms
in the Kubo bubble in Fig.\ref{fig:illustration}(a).
For these \emph{interband} processes,
the amplitude is determined by the
the probability of \emph{tunneling between two minibands upon measurement of velocity}.
We illustrate the relative contributions of interband and intraband processes to overall dc conductivity on Fig.~\ref{fig:illustration}(d) in 5 different regions of parameters of the FLA toy model.
These plots reveal that the diagonal components
of the velocity
cannot account for the regular sinusoidal BZ
oscillations, but only for the fractal behavior that is observed at
low $\Gamma$. It is
interesting to note that even at very high $\Gamma$, the intraband
processes still exhibit strong fractal behavior, while the overall
conductivity is already devoid of any apparent QOs. 
This indicates that the regular BZ oscillations are not a simple
``smoothing'' of the fractal behavior due to widened peaks in the
(fractal) spectral function. Rather, this is a separate phenomenon, ultimately due to oscillations in the tunneling amplitudes $v_{\tilde{\mathbf{k}},m,m'\neq m}$.

\emph{Relation to experiment.}
Both the fractal behavior (peaks in $\sigma_\mathrm{dc}^{xx}$ up to $p/q=4/q$) and the regular BZ oscillations have been observed in experiment\cite{KrishnaKumar2017,KrishnaKumar2018}.
The $T$-trend observed in Fig.~\ref{fig1}(d,e,f) is in qualitative agreement with the experimental findings of Ref.~\onlinecite{KrishnaKumar2017}.
Note that the lattice in this moir\'e system is different from that in our model, and that
the dominant interaction in graphene at high $T$ is likely of the
electron-phonon (e-ph) type,
while our Hamiltonian only includes e-e repulsion. The agreement
despite such differences indicates a significant level of universality in these phenomena.
Notwithstanding, the doping trend at the highest temperature is in apparent contrast to the measurements in Ref.~\onlinecite{KrishnaKumar2017}.
In our Fig.~\ref{fig1}(f), BZ oscillations are regular (sinusoidal) close to half-filling;
closer to the empty band limit a stronger fractal behavior remains in place.
In the corresponding high-$T$ experimental result in Ref.~\onlinecite{KrishnaKumar2017} (Fig.2B,C), only the regular oscillations are observed, 
and no oscillations at all are observed close to the ``neutrality point'' (corresponding to the empty band limit in our calculations).
This discrepancy appears to be due to the difference in the scattering mechanism: the e-e scattering rate goes to zero as the band empties, but the e-ph scattering rate does not.
The FLA calculation\cite{OURPRB} where $\Gamma$ is fixed regardless of the doping
clearly reproduces the doping-trend observed in the experiment.
Similarly, at low temperature in the Hubbard model, one observes both the SdH oscillations and fractal behavior (Fig.~\ref{fig1}(b)); In experiment, there are cases where only SdH oscillations are observed at low temperature. This discrepancy is, again, likely due to the difference in scattering mechanisms.
In the Hubbard model the scattering rate goes down with temperature (Fig.\ref{phaseDiagram}(c)). 
If the scattering rate is kept fixed at a moderate value (as in FLA), at low $T$ one only observes the SdH effect (see bottom panel in Fig.~\ref{fig:illustration}(d)).



\emph{Conclusion.} 
We have studied the magnetic quantum oscillations of longitudinal DC conductivity in the 2D Hubbard model.
We observe three types of non-monotonic behavior in $\sigma^{xx}_\mathrm{dc}$: 
1) Shubnikov-de Haas oscillations with frequency $p/q=n_\sigma$ (and higher harmonics), at low temperature;
2) fractal behavior of conductivity with peaks at $\Phi/\Phi_0=1/q,2/q,3/q...$, in the coherent regimes;
3) sinusoidal $p/q=1$-frequency oscillations, in moderately incoherent regimes (the Brown-Zak oscillations, BZ).
Our findings are in striking agreement with recent experimens on graphene superlattices.
The discrepancies from experiment can be traced back to a difference in interactions present in the system.
The oscillation phenomenology crucially depends on the scattering rate,
and can thus be used in experiment as a characterization tool for scattering mechanisms.
The fractal behavior is ultimately a manifestation of the Hofstadter butterfly, and is an indication of a low scattering rate;
in contrast, the BZ oscillations indicate a higher scattering rate, and when observed at very low temperature are an indication of a strong e-e coupling.
Our results present clear predictions for future experiments where the dependence on coupling strength and doping might be investigated.

\begin{acknowledgments}
Computations were performed on the PARADOX su- percomputing facility
(Scientific Computing Laboratory, Center for the Study of Complex
Systems, Institute of Physics Belgrade). J. V. acknowledges funding
provided by the Institute of Physics Belgrade, through the grant by
the Ministry of Education, Science, and Technological Development of
the Republic of Serbia, as well as by the Science Fund of the Republic
of Serbia, under the Key2SM project (PROMIS program, Grant No.
6066160). R. \v{Z}. is supported by the Slovenian Research Agency
(ARRS) under Program P1-0044 and Projects J1-1696 and J1-2458.
\end{acknowledgments}

\bibliography{paper1,synthetic}
\bibliographystyle{apsrev4-1}
\end{document}